\documentclass[aps,10pt,pra,twocolumn,groupedaddress, floatfix, superscriptaddress,showpacs, showkeys,amsfonts]{revtex4-2}
\usepackage{amsfonts}
\usepackage{amsmath}
\usepackage{amssymb, setspace}
\usepackage{graphics,graphicx}
\graphicspath{{./}}
\usepackage{epsf}
\usepackage{subfigure}
\usepackage[%
  colorlinks=true,
  urlcolor=blue,
  linkcolor=blue,
  citecolor=blue
]{hyperref}
\usepackage[all]{hypcap}
\usepackage{color}
\usepackage[utf8]{inputenc}
\usepackage{dsfont}
\usepackage{enumitem}
\usepackage[utf8]{inputenc}

\usepackage{comment}
\usepackage{orcidlink}
\usepackage{svg}

\usepackage{natbib}
\usepackage{braket}
\usepackage{physics}
\usepackage{gensymb}
\newcommand{\be}{\begin{equation}}
\newcommand{\ee}{\end{equation}}
\newcommand{\bea}{\begin{eqnarray}}
\newcommand{\eea}{\end{eqnarray}}

\newcommand{\fig}[1]{Fig.~\ref{#1}}

\newcommand{\bem}{\begin{multline}}
\newcommand{\eem}{\end{multline}}

\newcommand\identity{1\kern-0.25em\text{l}}

\begin{document}

\title{Demonstration of a partial knowledge quantum key distribution protocol}
\author{Liran Shirizly}
\altaffiliation{These authors contributed equally to this work.}

\author{Leonid Vidro}
\altaffiliation{These authors contributed equally to this work.}

\author{Hagai S. Eisenberg}
\email{hagai.eisenberg@mail.huji.ac.il}
 \affiliation{Racah Institute of Physics, Hebrew University of Jerusalem,
Jerusalem 91904, Israel}

\begin{abstract}
We demonstrate a novel QKD protocol, named \texttt{Bridge[$d,q$]}, which generalizes the \texttt{SARG04} protocol to $d$ dimensions and $q$ bases. 
In this protocol, Alice and Bob gain only partial knowledge about each other's raw key.
Nevertheless, this partial information is sufficient to distill a definite mutual encryption key.
We present an experimental demonstration of the \texttt{Bridge[$2,3$]} protocol. The three unbiased polarization bases were switched using a single Pockels cell.
The applied polarization basis switching method is relevant to any 3-basis QKD protocol and to other applications that require fast manipulation of photon polarization qubits.
\end{abstract}

\maketitle

\pagebreak
\section{\label{sec:Intro}Introduction}
Since the introduction of the \texttt{BB84} protocol~\cite{BB84,6states}, in most protocols for quantum key distribution (QKD), the communicating parties acquire full information about their measured states. However, in some situations, a quantum measurement does not provide full information about the prepared state but only partial information. Still, it is of interest to extract useful information in these situations.

One such example is when a measurement is performed using the Hong-Ou-Mandel effect (HOM)~\cite{HOM,HOM_timebins}. In this case, the analyzed photon and a known test photon are synchronized to arrive simultaneously at the two ports of a balanced beamsplitter. When the photons are identical, they bunch.

In a quantum state tomography scenario with many repetitions, the probability of bunching quantifies the projected overlap between the two photons' states. However, in a single-shot scenario, as in QKD, bunching provides no information about the photons' state overlap.

However, if the photons end at opposite detectors, they must carry different states, or they would have bunched. In more than two dimensions, this gives partial information on the unknown photon state - a set of states to which the photon does not belong.

Another way to explore the case of partial information is to extend the \texttt{SARG04} protocol~\cite{SARG04,SargDemo}, in which Alice and Bob announce their bits rather than their bases. This situation is identical to the one described above. When the received bit is identical to the one that was sent, no information is obtained. However, when the two bits differ, the bases must have been different too. Since the \texttt{SARG04} protocol utilizes only two bases, Alice and Bob still gain full shared information.

We introduce the \texttt{Bridge[$d,q$]} protocol (the name is inspired by the banned Highly Artificial Bidding Systems in the Bridge card game), which generalizes this situation to $d$-dimensions and $q$-bases. In such cases, each successful sifting event provides Alice and Bob with only partial information about their respective bases.

In this paper, we experimentally demonstrate the \texttt{Bridge[2,3]} protocol using polarization-encoded photons in three mutually unbiased bases. In addition to examining the newly proposed protocol, we use a single Pockels cell to randomly select one of the three bases at a high rate~\cite{3PockelsKwiat,3PockelsNIST}. Thus, the number of active elements is reduced. This approach can also be useful for other applications in quantum optics that require the use of the 3 unbiased bases with a single control element, such as quantum computation~\cite{Oneway,MBQC}.

The structure of this work is as follows. In Sec. \ref{sec:Protocol}, the \texttt{Bridge} protocol is presented. Sections \ref{sec:Setup} and \ref{sec:Results} present the experimental setup and the results, respectively. We conclude with a summary and discussion in Sec. \ref{sec:Summary}.

\section{\label{sec:Protocol}The Bridge Protocol}

In the new \texttt{Bridge} protocol, the two communicating parties, Alice and Bob, should have different raw outcomes after sifting, even under perfect transmission conditions. This is a major difference from other QKD protocols, where the raw keys in a noiseless scenario coincide.
  
In the \texttt{SARG04} protocol, Alice and Bob
define two mutually unbiased bases (MUBs)~\cite{MUB_prime}: 
\begin{equation}\label{MUB-2d}
    \begin{split}    
    \mathcal{B}_{z} & =\left\{ \left|\psi_{\pm}^{z}\right\rangle \right\} =\left\{ \left|0\right\rangle ,\left|1\right\rangle \right\} \\
    \mathcal{B}_{x} & =\left\{ \left|\psi_{\pm}^{x}\right\rangle \right\} =\left\{ \frac{1}{\sqrt{2}}\left(\ket{0}+\ket{1}\right), \frac{1}{\sqrt{2}}\left(\ket{0}-\ket{1}\right) \right\}\,. \\
    \end{split}
\end{equation}
For each photon $i$, Alice picks a basis bit $b^A_i \in \left\{ x,z\right\}$ and a state bit $s^A_i \in \left\{ +,- \right\}$. She encodes the photon in a random state $\left|\psi_{s^A_i}^{b^A_i}\right\rangle$ and sends it to Bob over a quantum channel. Bob chooses a random basis $b_i^B$ in which to measure the photon, yielding an outcome $s_i^B$. After each bit was measured, they reveal their chosen state bits, keeping only the cases $s_i^A\neq s_i^B$, where those state bits are different. In those sifted cases, each of their respective basis bits should also be different $b_i^A\neq b_i^B$, assuming no noise or eavesdropping. Since there are only two basis options, applying a \texttt{NOT} gate to one of the basis bits will result in identical bits for the raw key.\\
\\
\\
The simplest generalization, the \texttt{Bridge[2,3]} protocol, is for $d=2$ dimensions and $q=3$ bases. For a two-dimensional Hilbert space, there exists a third MUB in addition to those of Eq. \ref{MUB-2d}
\begin{equation}\label{MUB-3q}
    \begin{split}    
    \mathcal{B}_{y} & =\left\{ \left|\psi_{\pm}^{y}\right\rangle \right\} =\left\{ \frac{\left|0\right\rangle + i\left|1\right\rangle }{\sqrt{2}},\frac{i\left|0\right\rangle + \left|1\right\rangle }{\sqrt{2}} \right\}\,.   
    \end{split}
\end{equation}
In this case, both Alice and Bob choose a basis from a set of three options $b^A_i,b^B_i \in \left\{ x,y,z\right\}$. In the sifting stage, when they only keep results where $s_i^A\neq s_i^B$, for which $b_i^A\neq b_i^B$, they don't get full information, as the last condition is satisfied by two options. For example, if Alice encoded a state in the $\mathcal{B}_{x}$ basis, and Bob chose to measure in the $\mathcal{B}_{y}$ basis, after sifting Alice knows that Bob did not measure in the $\mathcal{B}_{x}$ basis, but cannot decide between the $\mathcal{B}_{y}$ and the $\mathcal{B}_{z}$ bases. The same situation exists for Bob, as he cannot decide between the $\mathcal{B}_{x}$ and the $\mathcal{B}_{z}$ bases. Each party is thus left with a conditional entropy of 1 bit of unknown information about the other's basis. Since the overall encoded information is one trit, they now share a mutual information of $\log_2 3 - 1 \approx 0.585$ bits ($\approx 0.37$ trits), which sets the upper bound for the noiseless key-rate.

To distill a concrete mutual raw key from this mutual information, they need only to exchange classical information. Alice has a sequence $\left\{t^A_i \right\}_{i=1}^N$ of $N$ classical trits $t^A_i \in \left\{ 0,1,2\right\}$, corresponding respectively to the bases $b^A_i \in \left\{ x,y,z\right\}$. Bob has his own sequence $\left\{ t^B_i \right\}_{i=1}^N$ , such that after sifting $t^B_i \neq t^A_i$. They need to map their sequences of partly anti-correlated outcomes to an identical shared sequence of bits. Efficient information reconciliation protocols that saturate the information-theoretic bound are beyond the scope of this work, and will be presented in Ref.~\cite{BridgeTBD}. Nevertheless, for completeness, we present a naive inefficient method here. For each index $i$, Alice randomly sends to Bob one of the two trit values that she \textbf{does not} have. With probability 0.5, it is not his trit, and by elimination he concludes that her trit is the third option. Otherwise, they neglect this trit transmission. On average, they are left with half a trit, while an eavesdropper obtains half of the difference between a trit and a bit. The overall rate of this reconciliation protocol is 0.5 bits per transmission.

We denote by \texttt{Bridge}[$d,q$] the general protocol for any $d$ dimensions and $q$ MUBs. In that case, there are $q$ bases $\mathcal{B}_{i} \in \left\{ \mathcal{B}_{1},\mathcal{B}_{2},...,\mathcal{B}_{q} \right\}$ and $d$ possible states for each basis $\left\{\left|\psi_{1}^{\mathcal{B}_{i}}\right\rangle,\left|\psi_{2}^{\mathcal{B}_{i}}\right\rangle,...,\left|\psi_{d}^{\mathcal{B}_{i}}\right\rangle \right\}$. The \texttt{SARG04} protocol is the special case of \texttt{Bridge[2,2]}.

\begin{figure*}
    \centering
    \includegraphics[width= 1\textwidth]{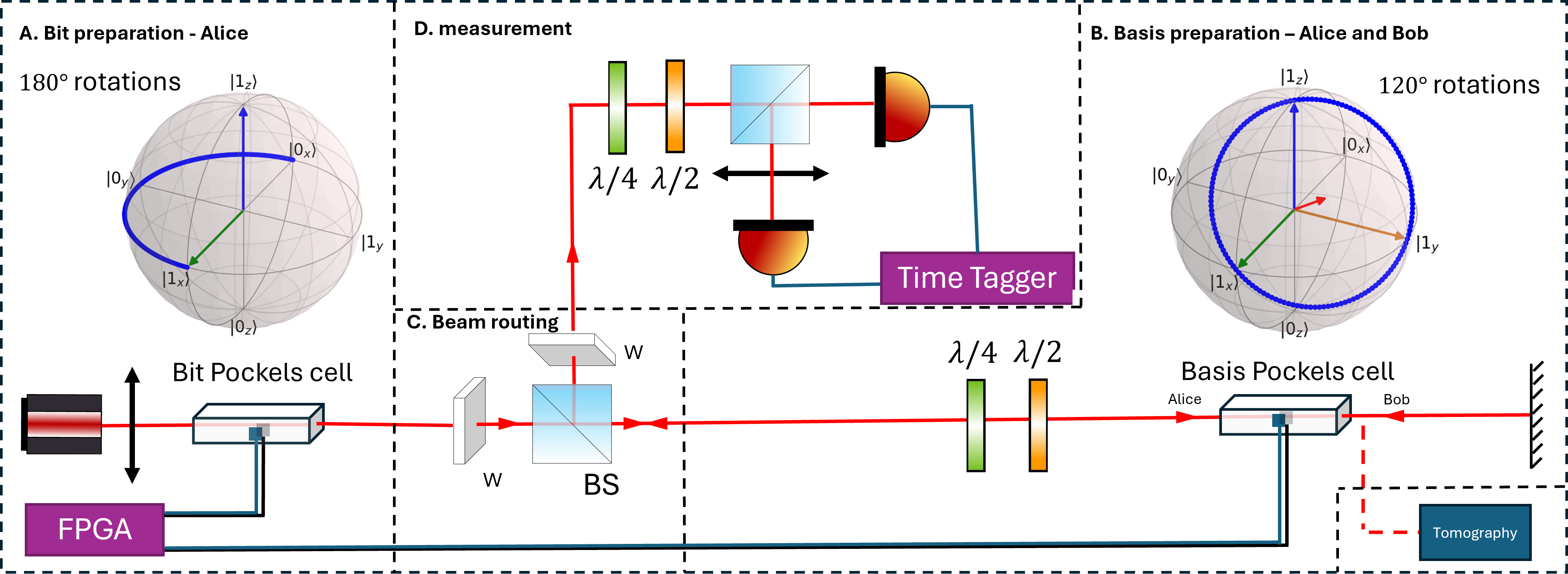}
    \caption{\textbf{The experimental setup. A.} The photon source is an attenuated Ti:Sapphire laser at $780$\,nm and $76$\,MHz reprate. A polarizer initializes the state, and the bit Pockels cell defines Alice's state bit $s^A$. \textbf{B.} A single Pockels cell chooses between one of 3 bases by rotation of $0,\pm2\pi/3$ on the Bloch sphere. Two waveplates align the polarization with respect to the Pockels cell's natural axes to achieve the correct rotation. The photons traverse back and forth along this path, rotated twice by the same Pockels cell, but with different rotations in each direction, defining different bases for Alice and Bob. \textbf{C.} The beam splitter serves the function of a circulator, taking the beam first from the bit choice to the basis choice, then from the basis choice to the projection stage. Two windows were used to compensate for the beamsplitter's polarization dependence by exploiting the polarization dependence of their Fresnel coefficients. \textbf{D.} A PBS projects the polarization to two detectors and an additional polarizer is used at the reflection arm to maximize the extinction ratio. Two waveplates align the polarization axes to those of the PBS such that when Alice and Bob choose the same basis, the axes of the initialization polarizer and the PBS coincide. An FPGA controls the low-voltage modulation signals to the Pockels cells and also generates a pseudo-random choice of bits and bases for Alice and Bob. A time tagger records the measurements. When polarization tomography is performed to assess the quality of the rotations of the Basis Pockels cell, the beam is sampled after a single pass through the Pockels cell.} 
    \label{fig:setup}
\end{figure*}

\section{\label{sec:Setup}Experimental Setup}

The experimental setup is illustrated in  Fig.~\ref{fig:setup}. We used a weak coherent photon source with $\mu\approx0.13$ mean number of photons per pulse, attenuated from a Ti-sapphire laser at 780\,nm, with a repetition rate of $76$\,MHz.
Alice's bit preparation consists of a polarizer and a Pockels cell ('bit Pockels cell') to switch between two orthogonal polarization states, choosing Alice's bit $s^A \in \left\{ +,- \right\}$.
For basis preparation, a second Pockels cell ('basis Pockels cell') was used to switch between three orthogonal directions on the Bloch sphere (three unbiased bases), choosing Alice's basis $\mathcal{B}_{b^A_i}$. To simplify the demonstration setup, the same Pockels cell was also used to set Bob's choice of Basis $\mathcal{B}_{b^B_i}$ as the photons traversed in the opposite direction.
A beam splitter was used between the two Pockels cells to separate the incoming and returning beams. Consequently, a portion of the beam intensity that exits the bit Pockels cell and another portion of the beam reflected towards the measurement setup are lost~\footnote{It has no effect on the performance of the setup since the power of the weak coherent source can be adjusted to suit the desired conditions of the experiment, accounting for the overall losses in the setup.}.
Two tilted uncoated glass windows were used to compensate for the beam splitter's polarization imbalance by exploiting the imbalance in the windows' Fresnel transmission coefficients.
Before measurement, a pair of a $\frac{\lambda}{4}$ and a $\frac{\lambda}{2}$ wave plates was used to adjust one of the polarization bases to coincide with the measurement axis defined by the PBS. An additional polarizer was placed at the PBS's reflection port to improve the extinction ratio of the reflected projection.

For the purpose of using the basis Pockels cell twice, a mirror after the Pockels cell reflected the light back for another passage. The distance that the light traverses from the Pockels cell to the mirror and back should provide enough time for the Pockels cell to change state to serve both Alice's and Bob's functionalities. After reflection, the polarization is flipped about the vertical axis, which is equivalent to a $\pi$ rotation about the Z-axis of the Bloch sphere (diagonal polarization goes to anti-diagonal polarization, right-hand to left-hand). However, the major axes of the Pockels cell are flipped as well, so that the Pockels cell continues to rotate the polarization of the back-reflected light in the same direction as the incoming light. For instance, if the Pockels cell is set to function as a quarter-wave plate, a photon that went through the Pockels cell forward and back would experience a rotation of a half-wave plate.
A different approach would be to place a quarter-wave plate at $0\degree$, which serves as an overall half-wave plate that flips the polarization once more. In this case, the Pockels cell effectively rotates the polarization back upon reflection, with an overall residual rotation.
It is important that the reflecting mirror will not add a phase. Thus, a metal mirror was used at an incident angle of $0\degree$ so that there would be as little difference as possible between TE and TM polarization modes.

An important feature of the setup is the ability to use a single Pockels cell to select between three unbiased bases. It is possible to rotate a state $s^A_i$ between the three unbiased bases $b^A_i$ on the Bloch sphere if the axis and angle of rotation are chosen correctly. Consider a trirectangular tetrahedron~\footnote{A tetrahedron being a triangular pyramid and a trirectangular tetrahedron being a tetrahedron where all three face angles at one vertex, which is the apex, are right angles}, with the origin of the Bloch sphere as the apex and the positive part of the three axes as edges going out from it. The height of the pyramid forms an angle $\theta_{t} = \arccos (\frac{1}{\sqrt{3}}) \approx 54.7\degree$ with the axes, and a rotation about the height of the pyramid by $2\pi/3$ takes each edge/axis to its neighboring edge/axis. The rotations are illustrated in Fig.~\ref{fig:setup}B.

A wave-plate or a Pockels cell, which is a fast and adjustable wave-plate, can add a phase between two orthogonal linear polarization components, depending on the orientation of its major axes. Thus, the natural axis of rotation of the Pockels cell is on the $X-Z$ plane of the Bloch sphere and cannot be used directly to rotate between the standard 3 axes (bases). However, it can be used to choose from a different triad of unbiased bases that can be rotated to any other triad later. A pair of $\frac{\lambda}{4},\frac{\lambda}{2}$ wave plates is used before the basis Pockels cell to adjust the input polarization to the angle $\theta_{t}$ with respect to the rotation axis of the Pockels cell, to achieve the desired "tetrahedron" rotation. This alignment also accounts for the crystal's natural birefringence.

The polarization rotation of a Pockels cell is set by the voltage difference between its two electrodes. The voltage on the basis Pockels cell could be set and controlled (On/Off) separately for each electrode, allowing switching between four different polarization rotations.
The two high voltage values of the basis Pockels cell were both set to $2V_{\pi}/3$, where $V_{\pi}$ denotes the voltage required for a rotation of $\pi$ on the Bloch sphere. This setting enables the four rotations: $0,-2\pi/3,2\pi/3,(2\pi/3-2\pi/3)\approx0$ (turning both voltages 'On' is equivalent to keeping both 'Off' since the voltage difference across the Pockels cell in this case is $0$). A different configuration of the two high voltages, $2V_{\pi}/3$, and $4V_{\pi}/3$, could also be used in a similar way.

The polarization measurement is performed by a PBS. When Alice and Bob choose the same basis, the photons' polarization should be aligned with the projection axis. If they choose different bases, the exact axis to which the photons are aligned should not be important. The sent state basis just has to be mutually unbiased with the measurement basis (or orthogonal on the Bloch sphere).
When Alice and Bob choose, respectively, to rotate either $0\degree$ and $0\degree$ or $2\pi/3$ and $-2\pi/3$ or $-2\pi/3$ and $2\pi/3$, the output polarization basis is the same as the input. Those three cases represent the events in which Alice and Bob chose one of the three options for the same basis. Other choices of Alice and Bob result in the preparation and measurement of the polarization along orthogonal axes on the Bloch sphere.

The bit Pockels cell is a \textit{Conoptics} transverse field \textit{KD*P} crystal Pockels cell with configurable bias, an adjustable 'On' voltage, and 30\,MHz max switching rate.
The basis Pockels cell is an X-cut \textit{RTP} crystal by \textit{Leysop} with a 3\,ns rise time and 10\,MHz max switching rate driver by \textit{BME}.
The photons were detected by Avalanche Photodiodes (APDs).
The Pockels cells were modulated by a low-voltage signal from an FPGA, which generated pseudo-random sequences for the bits and the bases, and controlled the timing of the modulation signal.

The Modulation frequency was adjusted to $0.9$\,MHz to minimize acoustic ringing in both Pockels cells~\cite{Pockels_cell_vaib}. 
In this case, the system changes its state every 84 laser clock cycles. The modulation signal to the basis Pockels cell was timed to change its state exactly when one of the photons exits the Pockels cell. By the time this photon is reflected back to the basis Pockels cell, its state is changed. Thus, one of every 84 photons has a different basis choice for Alice and Bob and is used to demonstrate the protocol.
The detection signals were registered using a time-tagger and analyzed offline.

\begin{figure}
    \centering
    \includegraphics[width=1\linewidth]{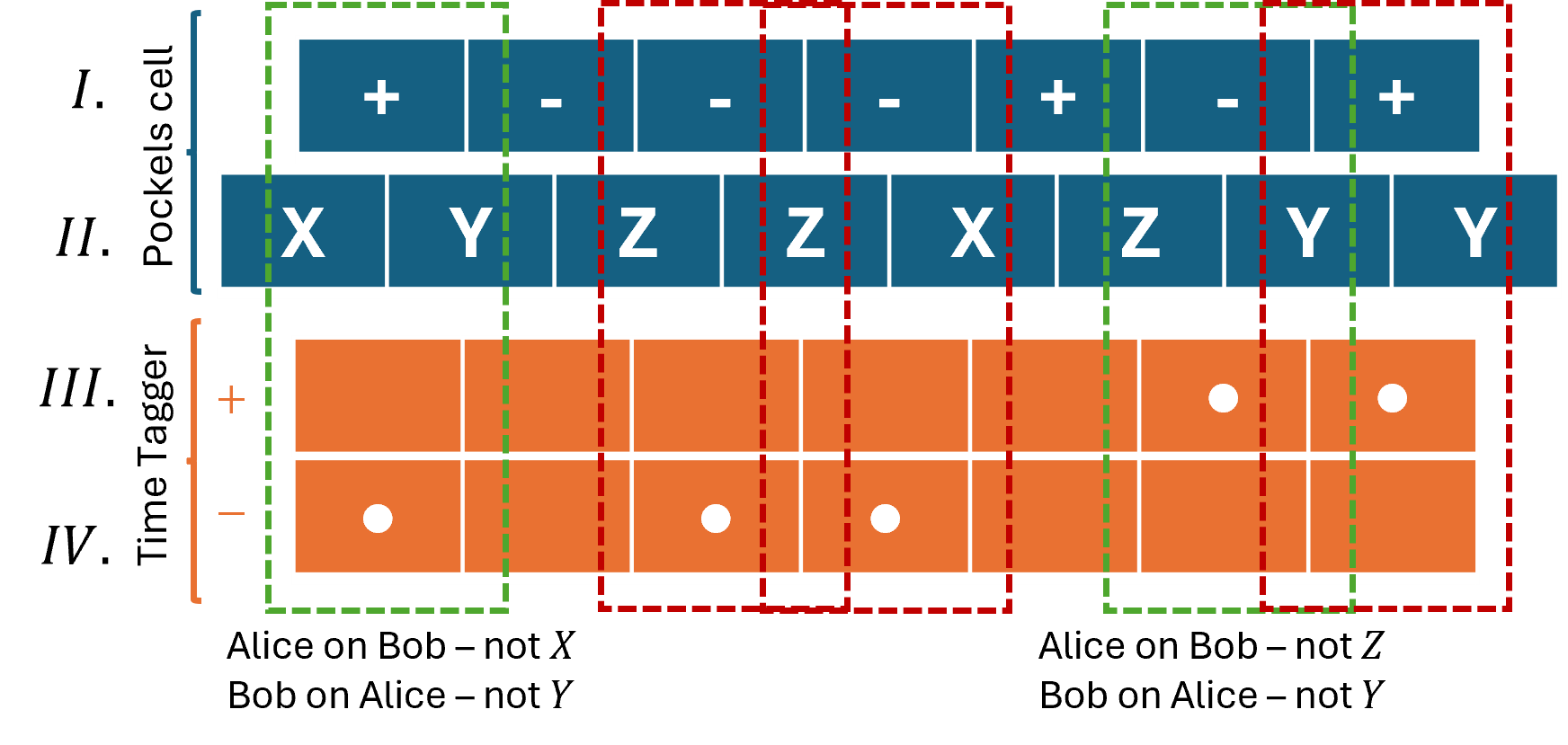}
    \caption{A typical sequence of modulation, detection, and sifting. Modulation sequences created by the FPGA are in blue, detection sequences received by the time-tagger are in orange. Alice's sequences of bits $s^A\in\{+,-\}$ are in row $I.$ The sequence of Bases, $b_i^A,b_i^B\in\{X,Y,Z\}$, is in row $II.$ (Both for Alice and for Bob because of the double function of the basis Pockels cell). A white circle represents a photon detected (if detected) in one of Bob's detectors, resulting in a bit $s_B\in\{+,-\}$ in rows $III.$ and $IV.$ accordingly. A green dashed box represents events that passed sifting - where Bob and Alice had opposite bits. A red dashed box represents events that didn't pass sifting. After sifting, both sides are left with partial information.}
    \label{fig:sifting}
\end{figure}

\section{\label{sec:Results}Results}
\begin{figure}[t]
    \centering
    \includegraphics[width=0.35\textwidth]{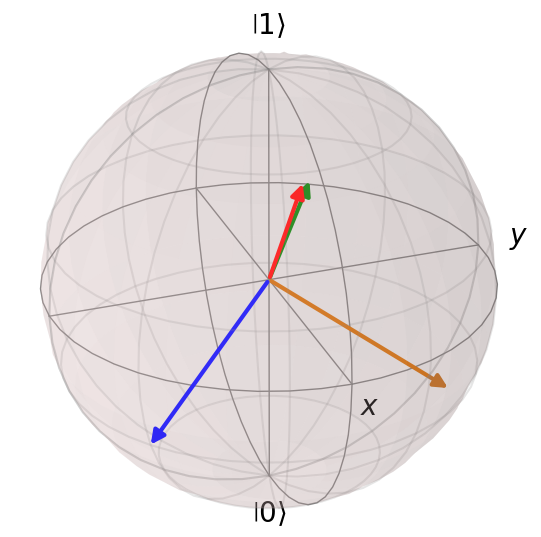}
    \caption{Experimental polarization state tomography of Alice's $s^A=+$ prepared state after the rotation of the Basis Pockels cell. The green, blue, orange, and red correspond to the 4 voltage levels, which generate rotations around the x-axis by $0\degree$, $2\pi/3$, $-2\pi/3$, and $(2\pi/3-2\pi/3)$ respectively. The experimental mean angle between the states is $88.6\pm1.8\degree$.}
    \label{fig:Alice_states}
\end{figure}

A QKD protocol requires high fidelity for state preparation and measurement, and a low communication-line error rate. In order to characterize the state preparation and the rotation to the three MUBs, we applied a quantum state tomography (QST) protocol~\cite{QST} to the photon state after a single pass through the basis Pockels cell, as shown in the bottom right corner of Fig.~\ref{fig:setup}.

The state was initially set to $\left|\psi_+^{z}\right\rangle$ by a polarizer and the bit Pockels cell.  The Pockels cell axis of polarization rotation was the Bloch $x$-axis. Following this, the state's polarization was rotated by the first set of wave plates to an angle of $54.73\degree$ from the $x$-axis in the $X-Z$ plane. A quantum state tomography protocol characterized each of the four possible rotated states. The measured Bloch vectors are shown in \fig{fig:Alice_states}. The average angle between the orthogonal axes is $88.6\pm1.8^\circ $.  

The synchronization of the measured data with the modulation sequences applied to the Pockels cells was performed offline using timestamps received from the time-tagger. Figure~\ref{fig:sifting} presents an example of a typical received sequence of detections by Bob, synchronized with the modulation sequences created by the Bit Pockels cell to choose Alice's bit and the Basis Pockels cell to choose the basis for both Alice and Bob. For the purpose of the demonstration, we can choose any adjacent pair of Bases and treat the first one as Alice's and the second one as Bob's. After the sifting process, Alice and Bob each retain partial information about the other's basis. With a further information reconciliation procedure, a common key is formed.

\begin{figure}
    \centering\includegraphics[width=1\linewidth]{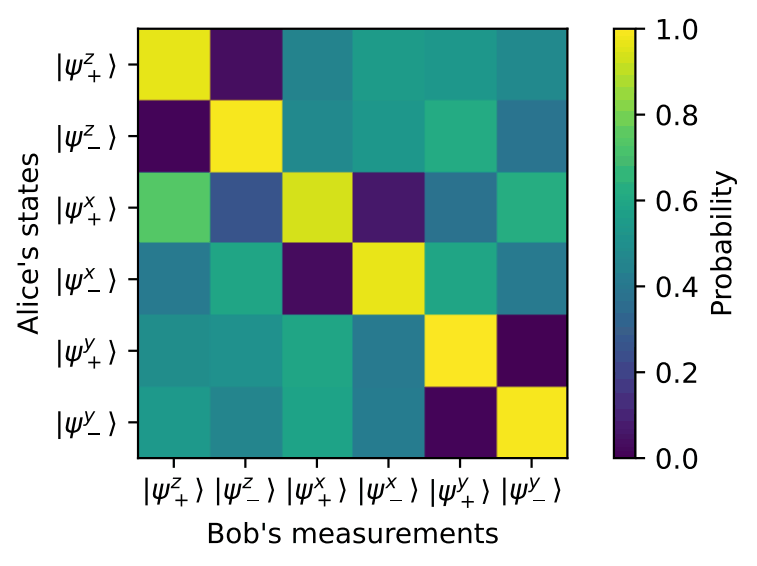}
    \caption{Statistics from the protocol run of the detection probabilities of all combinations of Alice's prepared states $\left|\psi_{s^A}^{b^A}\right\rangle$ and Bob's measurement outcomes, depending on his choice of basis,  $\left|\psi_{s^B}^{b^B}\right\rangle$. On the diagonal 2X2 blocks, Alice and Bob chose the same basis, and thus, the result is close to the identity matrix. On the off-diagonal blocks, the bases are unbiased, and all entries are close to $\frac{1}{2}$. Actually, each row is a full tomography measurement of one of Alice's states with an average fidelity of $0.97\pm0.02$.}
    \label{fig:matrix}
\end{figure}

To characterize the quality of the entire setup, we use the same collected data, both sifted and unsifted. Statistics of Bob's outcomes $s^B_i \in \left\{ +,- \right\}$ is measured for all possible combinations of Alice's bit $s^A_i \in \left\{ +,- \right\}$ and preparation basis $b^A_i\in \left\{ x,y,z \right\}$, as well as for Bob's measurement basis $b^B_i \in \left\{ x,y,z \right\}$. A total of 36 combinations are presented in Fig.~\ref{fig:matrix}. These are measurements taken during the protocol's operation, analyzed statistically rather than used to create a key. Thus, they reflect the quality of the quantum link. If Alice and Bob choose the same basis, we expect to clearly see the identity sub-matrix, represented on the diagonal as 2-by-2 blocks. When they choose different bases, we expect a sub-matrix with unbiased, equal probabilities of $\frac{1}{2}$. This can be seen for all non-diagonal blocks with values of $0.5\pm 0.1$. Actually, the results of each row are the measurements required for a state tomography of one of Alice's sent states. The average fidelity of this tomography is $0.97\pm0.02$.

The rate of any QKD protocol depends on the quantum bit-error-rate (QBER) of the overall system. This rate is the result of calculations used to prove the protocol's security. The QBER $Q$ is defined as the ratio of the number of incorrect outcomes to the total number of transmissions that passed sifting. In this protocol, incorrect outcomes occur when Alice and Bob use the same basis but measure different bits. Such cases would pass sifting, wrongly suggesting that they have different bases, with probability $\text{Pr}(s^A\neq s^B|b^A=b^B) \text{Pr}(b^A=b^B)$. Unlike \texttt{BB84}, where the sifting preserves matching bases, here the sifting depends on the measurement outcomes. Therefore, channel noise changes the sifting rate. The probability of passing sifting is the sum of all the possibilities for different measurement outcomes 
\begin{equation}
\begin{aligned}
    p_{\text {pass}} =& \text{Pr}(s^A \neq s^B|b^A \neq b^B) \text{Pr}(b^A \neq b^B)  \\ +&
    \text{Pr}(s^A\neq s^B|b^A=b^B) \text{Pr}(b^A=b^B)
\end{aligned}
\end{equation} 
 Using uniform basis selection, so that $\text{Pr}(b^A\neq b^B)=2\,\text{Pr}(b^A=b^B)$, the basis-choice probabilities cancel and the error rate is
\be\label{Eq:error_def}
    Q = \frac{\text{Pr}(s^A\neq s^B|b^A=b^B)}{2\text{Pr}(s^A\neq s^B|b^A \neq b^B) + \text{Pr}(s^A\neq s^B|b^A=b^B)}  
\ee

\begin{figure}[t]
    \centering
    \includegraphics[width=0.47\textwidth]{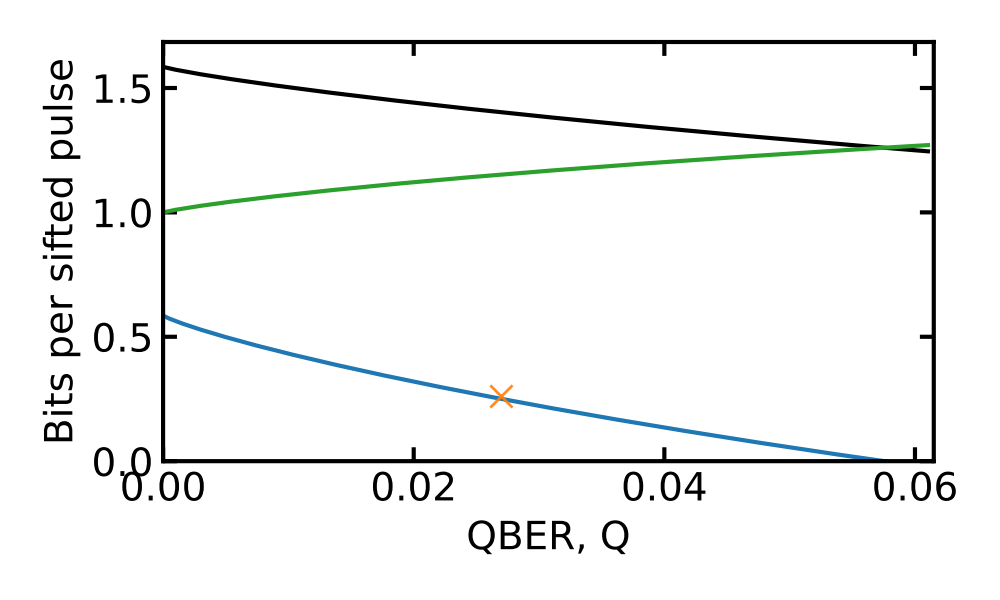}

    \caption{
    Secure key rate of the \texttt{Bridge}[$2,3$] protocol. The upper black line denotes the conditional entropy between Alice and Eve $H(A|E)$. In a noiseless system, this is 1 trit of information. The middle green line denotes the conditional entropy between Alice and Bob $H(A|B)$. In a noiseless scenario, this is one bit of information. The bottom blue line denotes the protocol rate calculated by the difference between the two conditional entropies.
    The protocol is viable up to an error rate of 5.75\%. 
    The finite conditional entropy of Alice on Bob, even at zero QBER, emphasizes the unique property of the protocol where Alice and Bob don't have complete information about each other's raw key.
    The 'X' mark corresponds to the measured error rate in our system.}
    \label{fig:key_rate}
\end{figure}

Figure~\ref{fig:key_rate} presents the secure key rate per sifted photon of the \texttt{Bridge}[$2,3$] protocol, under the assumptions of optimal information reconciliation, a symmetric depolarization channel, and a perfect single photon source. This result was obtained numerically using the framework of Ref. \cite{Norbet_reliablenumerical2018} (see also \cite{BridgeTBD}). In the absence of errors, the secure key rate is $\log_2 3 - 1\approx0.58$ bits. The critical QBER for extracting a secure key is 5.75\%.

In agreement with the numerical results, the conditional entropy of Alice on Bob can also be expressed analytically as
\be
H\left(A|B\right)=-Q\log_2 Q - \left(1-Q\right) \log_2 \frac{1-Q}{2},
\ee
which can be intuitively explained by the entropy of having the same or different bases with distribution $\left\{Q,\frac{1-Q}{2},\frac{1-Q}{2} \right\}$.

Considering the results presented in Fig.~\ref{fig:matrix}, the experimental average QBER is $Q=2.8\pm0.9 \%$. This is below the critical QBER of $5.75\%$, as shown with an 'X' mark in \fig{fig:key_rate}.
The conditional entropy between Alice and Bob is $H(A|B)=1.14\pm0.03$ bits per sifted photon, and the corresponding conditional entropy between Alice and Eve is $H(A|E)=1.41\pm0.04$. The secure key rate is therefore $H(A|E)-H(A|B)=0.26 \pm 0.04$ bits per sifted photon

The probability of receiving a photon from the attenuated laser pulse and detecting it is $2.5 \cdot 10^{-3}$ per laser pulse, due to losses. Since we use one every 84 laser pulses, and the repetition rate is 76\,MHz, there are $2.3\cdot10^3$ relevant detected photons per second. Given the secure key rate and $p_{\text {pass}}$, this proof-of-concept setup enables extraction of $200\pm20$ secure bits per second.

\section{\label{sec:Summary}Summary and Discussion}
In this work, we demonstrate the \texttt{Bridge}[$2,3$] QKD protocol. The Bridge protocol is motivated by scenarios in which a measurement gives only partial information about a quantum state. In this specific demonstration, where 3 bases are used and the information is encoded in the basis, a trit is being sent. In each sifting event, one of the three basis options is eliminated, resulting in partial information.

We also demonstrate a configuration in which a single Pockels cell rotates between 3 mutually unbiased bases. The states were created with high fidelity, both in a static configuration using separate tomography and in real time during the protocol's operation. This configuration can be useful also for other applications that require fast polarization rotation or even feed-forward, such as one-way quantum computation.

The maximum switching rate of the Pockels cell is 10 times higher than that used in the experiment. Thus, the protocol rate can be increased if acoustic ringing issues are resolved. Still, the switching rate of the polarization would be every 7-8 photons. Thus, the change in photons' polarization is not bit-by-bit, requiring the selection of relevant photons in a real scenario.

Further increasing the rate is possible by adjusting the mean photon number of the weak coherent state. The \texttt{SARG04} protocol is more robust against photon-number splitting attacks. However, the exact overall performance of the extended \texttt{Bridge} protocol is left for future work, and decoy states may be required~\cite{decoy1,decoy2}. Alternatively, a single photon source can be used.

Another extension of this demonstration is to use polarization-entangled photon pairs~\cite{Ekert91,BBM92} to obviate the need for a bit Pockels cell. The projection bases of both photons have to be chosen randomly, but the photons' bits are randomly projected by a PBS.
A further study of this protocol would involve higher-dimensional states, as in \cite{HOM_timebins}, where HOM measurements are used to characterize very short time-bin qudits.

\bibliography{bib.bib}

\end{document}